\documentclass[floats,prb,aps,twocolumn]{revtex4}
\usepackage{graphicx}
\begin{document}

\vspace{2.5cm} \preprint{{\em Submitted to Europhysics Letters}
\hspace{1.0in} Web galley}
\date{\today}

\title{Superfluidity of  ``dirty'' indirect excitons in coupled quantum wells}

\author{Oleg L. Berman$^{1}$, Yurii E. Lozovik$^{2}$,  David W.
Snoke$^{3}$, and Rob D. Coalson$^{1}$}

\affiliation{\mbox{$^{1}$ Department of Chemistry, University of
Pittsburgh,}   \\ Pittsburgh, PA 15260, USA  \\
\mbox{$^{2}$ Institute of Spectroscopy, Russian Academy of
Sciences,}  \\ 142190 Troitsk, Moscow Region, Russia \\
\mbox{$^{3}$Department of Physics and Astronomy, University of
Pittsburgh,}  \\ 3941 O'Hara Street, Pittsburgh, PA 15260 USA  }


\date{\today}


\begin{abstract}

The theory of what happens to a superfluid in a random field,
known as the ``dirty boson'' problem, directly relates to a real
experimental system presently under study by several groups,
namely excitons in coupled semiconductor quantum wells. We
consider the case of bosons in two dimensions in a random field,
when the random field can be large compared to the repulsive
exciton-exciton interaction energy, but is small compared to the
exciton binding energy. The interaction between excitons is taken
into account in the ladder approximation. The coherent potential
approximation allows us to derive the exciton Green's function for
a wide range of the random field strength, and in the
weak-scattering limit CPA results in the second-order Born
approximation. For quasi-two-dimensional excitonic systems, the
density of the superfluid component and the Kosterlitz-Thouless
temperature of the superfluid phase transition are obtained, and
are found to decrease as the random field increases.

\vspace{0.2 in}



\end{abstract}

\maketitle {}


Superfluidity in a system of spatially indirect excitons in
coupled quantum wells (CQW) has been predicted by Lozovik and
Yudson,\cite{Lozovik} and several subsequent
studies\cite{Lerner,Dzjubenko,Knox,
Yoshioka,Birman,Littlewood,Vignale,Berman_Tsvetus,DasSarma,Ulloa,Moskalenko}
have predicted that this should be manifested as persistent
electric currents, quasi-Josephson phenomena and unusual
properties in strong magnetic fields. In the past ten years, a
number of experimental studies have focused on this goal.
\cite{Snoke_paper,Snoke_paper_Sc,Chemla,Krivolapchuk,Timofeev,Zrenner,Sivan,Snoke}
The coupled quantum well system is conceptually simple: negative
electrons are trapped in a two-dimensional plane, while an equal
number of positive holes is trapped in a parallel plane a distance
$D$ away. One of the appeals of this system is that the electron
and hole wavefunctions have very little overlap, so that the
excitons can have very long lifetime  ($> 100$ ns), and therefore
they can be treated as metastable particles to which
quasiequilibrium statistics apply. Also, when $D$ is large enough,
the interactions between the excitons are entirely repulsive, so
that the exciton gas is stable against collapse (see, e.g.,
Ref.[\onlinecite{Berman,Berman_Willander}]).

In general, the theory of these systems has
 neglected the role of disorder. In real experiments, however,
 disorder plays a major role, arising from impurities, boundary
 irregularities, and fluctuations of the alloy concentration of the
 epitaxial layers. Although the quality of quantum well
structures has improved dramatically over the past decade, e.g.,
the inhomogeneous broadening linewidth of typical GaAs-based
samples has been improved from around 20 meV to less than 1
meV,\cite{Snoke_paper} the disorder energy is still not negligible
compared to other energies. At a typical exciton density of
$10^{10}$ cm$^{-2}$, the interaction energy of the excitons is
approximately $4\pi \hbar^2 a n/M D \simeq 1 $ meV ($a$ being the
characteristic radius of a $2D$ exciton, $M$ the total mass, and
$n$ the number density of excitons).\cite{Snoke} Typical thermal
energies at liquid helium temperatures are $k_BT = 0.2 - 2$ meV.
On the other hand, the typical disorder energy of 1 meV is low
compared to the typical exciton binding energy of 5 meV.

Earlier studies of disorder in exciton systems include theory of
the transport properties of direct and indirect excitons and
magnetoexcitons in random fields,\cite{Ruvinsky_jetp} the
influence of various random fields on excitonic and
magnetoexcitonic absorption of light,\cite{Ruvinsky_jetp} and
Anderson localization of excitons.\cite{Gevorkyan}   In
two-dimensional systems, the excitonic interaction in the
Bogoliubov approximation is valid only at non-physically low
densities because of the divergence of the two-dimensional
scattering amplitude in the Born approximation.\cite{Yudson}
Therefore, the ladder approximation must be used at low densities
to treat properly the interaction between two-dimensional
excitons.\cite{Yudson,Abrikosov}  In this letter we study the case
of a random field which is not necessarily small compared to the
dipole-dipole repulsion between excitons. The coherent potential
approximation\cite{Gevorkyan} (CPA) allows us to derive the 2D
indirect exciton Green's function for a wide range of the random
field, and in the weak-scattering limit CPA results in the
second-order Born approximation.

As mentioned above, fluctuations of the thickness of the quantum
wells, which arise during the fabrication process, impurities in
the system, and disorder in the alloy of the barriers can all lead
to the appearance of a random field. Of these, spectral analysis
of the exciton luminescence shows that alloy disorder, with a
characteristic length scale short compared to the excitonic Bohr
radius of around 100~\AA, plays the most important role. In our
model, the random potential $V(\mathbf{r})$ acting on an electron
and hole is considered to be a Gaussian noise, such that
\begin{eqnarray}
\label{gaussian} \left\langle V(\mathbf{r})V(\mathbf{r}')
\right\rangle = g \delta (\mathbf{r} - \mathbf{r}'),
\hspace{0.5in} \left\langle V(\mathbf{r}) \right\rangle = 0 .
\end{eqnarray}
 An electron is subjected to the potential $V_{e} = \alpha_{e}
V(\mathbf{r})$ and a hole to $V_{h} = - \alpha_{h} V(\mathbf{r})$,
where $\alpha_{e}$ and $\alpha_{h}$ are constants. We consider the
characteristic length of the random field potential  $l$ to be
much smaller than the average distance between excitons $r_{s}
\sim 1/\sqrt{\pi n}$  ($l \ll 1/\sqrt{\pi n}$, where $n$ is the
total exciton density). Therefore, in order to obtain the Green's
function of the excitons with dipole-dipole repulsion in the
random field,  we obtain the Green's function of a single exciton
in the random field  (not interacting with other excitons), and
then apply perturbation theory with respect to dipole-dipole
repulsion between excitons, using the gas of non-interacting
excitons as a reference system.

The momentum-frequency domain Green's function of the center of
mass of the isolated exciton in the random field at $T = 0$
 within the coherent potential approximation (CPA) is given by\cite{Gevorkyan} (here and below
$\hbar = 1$)
\begin{eqnarray}
\label{green_0} G^{(0)}(\mathbf{p}, \omega) = \frac{1}{\omega -
\varepsilon_{0}(p) + \mu + i Q(\mathbf{p}, \omega) }  ,
\end{eqnarray}
where $\mu$ is the chemical potential of the system, and
$\varepsilon_{0}(p) = p^{2}/2M$ is the spectrum of the center mass
of the exciton in the ``clean'' system ($M = m_{e} + m_{h}$ is the
mass of the exciton; $m_{e}$ and $m_{h}$ are the electron and hole
masses, correspondingly). The function $Q(\mathbf{p}, \omega)$ is
determined by the effective random field acting on the center of
mass of the exciton. For zero random field, $Q(\mathbf{p}, \omega)
\rightarrow 0$. If $g \ll E_{b}$ ($E_{b}$ is the binding energy of
an exciton), the function $Q(\mathbf{p}, \omega)$ in the
second-order Born approximation  can be expressed through the
ground-state wavefunction of an indirect exciton. Analytical
results for $Q(\mathbf{p}, \omega)$ may be derived for $D \gg a$
and $m_{e} = m_{h}$ or $m_{h} \gg m_{e}$ and $\alpha_{h} \left(
\frac{m_{h}}{m_{e}} \right)^{2} \ll \alpha_{e}$. We present in
this Letter the simplest case $m_{e} = m_{h}$.  The result, for
the case $D \gg a$ and $m_{e} = m_{h}$, is
\begin{eqnarray}
\label{Q_res}  Q (\mathbf{p}, \omega) &=& \frac{(\alpha_{e} -
\alpha_{h})^{2}M g}{16\pi^{4}} \exp\left( -\frac{\rho^{2}}{8}
(p^{2} + 2M\omega)\right) \nonumber\\
&& \times J_{0}\left( \frac{\rho^{2}}{4} \sqrt{2M\omega}p \right),
\end{eqnarray}
Here $\rho = (8a)^{1/4}D^{3/4}$, with $a = \epsilon/4m_{e-h}e^{2}$
($\epsilon$ is the dielectric constant of the material, $m_{e-h}$
is the reduced mass of exciton, and $e$ is the electron charge),
and $J_{0}(z)$ is a Bessel function of the first kind.

At small densities $n$ ($n \rho^2 \ll 1$), the system of indirect
excitons at low temperatures is a two-dimensional weakly nonideal
Bose gas of dipoles with dipole moments  ${\bf d}$ normal to wells
($d \sim eD$). The distinction between excitons and bosons
manifests itself in exchange effects (see, e.g.,
Refs.[\onlinecite{Berman,Berman_Willander}], and
[\onlinecite{Halperin,Keldysh}]). These effects are suppressed for
excitons with spatially separated $e$ and $h$ in a dilute system
($n\rho^2 \ll 1$)  at large $D$ ($D \gg a$),  because at large
$D$, the exchange interaction in the spatially separated system is
suppressed relative to the $e-h$ system in a single well due to
the smallness of the tunneling exponent  $T \sim
\exp[-(D/2a)^{1/4}]$ originating from the dipole-dipole
interaction. Hence, when $D \gg a$, exchange phenomena, connected
with the distinction between excitons and bosons, can be
neglected, and therefore, the system of indirect excitons in CQWs
can be treated by diagrammatic techniques developed for boson
systems\cite{Abrikosov}. Two indirect excitons in a dilute system
interact as $U(R) = e^{2}D^{2}/(\epsilon R^{3})$, where $R$ is the
distance between exciton dipoles along quantum well planes.

At the characteristic frequencies and momenta which give the
greatest contribution to the Green's function in the ladder
approximation, the function $Q (\mathbf{p}, \omega)$ can be
approximated by the constant (see Eq.~(\ref{Q_res})) $Q
(\mathbf{p} = \mathbf{0}, \omega = 0) =  Q = (\alpha_{e} -
\alpha_{h})^{2}M g/(16\pi^{4})$. As a result, the first order
approximation in the amplitude of scattering of the isolated pair
of bosons $f_{0}$ the two-particle vertex $\Gamma$ can be reduced
to $f_{0}$ pertinent to a clean system\cite{Abrikosov,Yudson},
i.e. $\Gamma$ does not depend on $Q$ in the first approximation.
As a result, the chemical potential $\mu$ has the same form as for
the pure system\cite{Yudson}:
\begin{equation}\label{Mu}
\mu =   \frac{8\pi n}{2M \log \left( \frac{\epsilon^{2}}{8\pi n
M^2 e^4 D^4} \right)} .
\end{equation}

We  have the condensate Green's function $D(\mathbf{p},
i\omega_{k})$
\begin{eqnarray}
\label{d0} D^{(0)}({\bf p},i\omega _{k}) = \frac{- i(2\pi )^{2}
n_{0}\delta ({\bf p})}{i\omega _{k} + \mu + i Q},
\end{eqnarray}
where $n_{0}$ is the density of Bose condensate. Since at small
temperatures $(n - n_{0})/n \ll 1$, according to the ladder
approximation\cite{Abrikosov} we use $n$ instead of $n_{0}$ below.
$G(\mathbf{p}, i\omega_{k})$ and $F(\mathbf{p}, i\omega_{k})$ are
the normal and anomalous Green functions of the overcondensate:
\begin{eqnarray}
\label{g0} G({\bf p},i\omega _{k}) &=& - \frac{i \omega _{k} +
\varepsilon _{0}(p)
   + \mu + i Q}{
\omega _{k}^{2} + \varepsilon ^{2}(p) - 2i (\mu - \varepsilon
_{0}(p))Q}  ; \nonumber\\
   F({\bf
p},i\omega _{k}) &=&  - \frac{\mu }{ \omega _{k}^{2} + \varepsilon
^{2}(p) - 2i(\mu - \varepsilon _{0}(p)) Q} ,
\end{eqnarray}
where $\varepsilon _{0}(p)$ is the spectrum of noninteracting
excitons; for small momenta $p \ll \mu$ the excitation spectrum of
the interacting excitons, $\varepsilon (p)$,  is acoustic:
$\varepsilon (p) = c_{s} p$, where $c_{s} = \sqrt{\sqrt{\mu^{2} -
Q^{2}}/M}$ is the velocity of sound.

We calculate the density of the normal component $n_{n}(T)$, using
the Kubo formula\cite{Mahan} and the total single-particle
Matsubara Green's function of the indirect excitons. We obtain
\begin{eqnarray}
\label{nn4} n_{n} = n_{n}^{0} + \frac{N}{M} \int_{}^{}\frac{d{\bf
p}}{(2\pi )^{2}} p^{2}\mu \frac{\varepsilon _{0}(p)}{ \varepsilon
^{4}(p)}Q       .
\end{eqnarray}
Here $N$ is the total number of particles, and $n_{n}^{0}$ is the
density of the normal component in a pure system with no
impurities:
\begin{eqnarray}
\label{Bose} n_{n}^{0}  = - \frac{1}{2M} \int_{}^{} \frac{d{\bf
p}}{(2\pi )^{2}} p^{2} \frac{\partial n_{0}(p)}{\partial
\varepsilon}.
\end{eqnarray}
where $n_{0}(p)=(e^{\varepsilon (p)/T} - 1)^{-1}$  is the
distribution of an ideal Bose gas of temperature excitations.

The first term in Eq.~(\ref{nn4}), which does not depend on $Q$, is the
contribution to the normal
component due to scattering of quasiparticles with an acoustic
spectrum in an ordered system at $T \neq 0$. In a two-dimensional system,
\begin{eqnarray}
\label{nn00} n_{n}^{0} =  \frac{3 \zeta (3) }{2 \pi }
\frac{T^3}{c_{s}^{4}(n,Q) M},
\end{eqnarray}
where $\zeta (z)$ is the Riemann zeta function ($\zeta (3) \simeq
1.202$).
   The second term in Eq.~(\ref{nn4})
is the contribution to the normal component due to the interaction
of the particles (excitons) with the random field, which, when
explicitly evaluated, yields:
\begin{eqnarray}
\label{nn55} n_{n} = n_{n}^{0} + \frac{n Q}{2Mc_{s}^{2}(n,Q)}  .
\end{eqnarray}
The density of the superfluid component is $n_{s} = n - n_{n}$.
  From Eqs.~(\ref{nn00}) and~(\ref{nn55}) we can see that increasing the
random field decreases the density of the superfluid component.

In a $2D$ system, superfluidity appears below the
Kosterlitz-Thouless transition temperature $T_{c} = \pi
n_{s}/(2M)$\cite{Kosterlitz}, where only coupled vortices are
present. Using the expressions Eqs.~(\ref{nn00}) and~(\ref{nn55})
for the density $n_{s}$ of the superfluid component, we obtain an
equation for the Kosterlitz-Thouless transition temperature
$T_{c}$. Its solution is
\begin{eqnarray}
\label{tct} T_c &=& \left[\left( 1 +
\sqrt{\frac{32}{27}\left(\frac{M T_{c}^{0}}{\pi n'}\right)^{3} +
1} \right)^{1/3}  \right. \nonumber \\
&-& \left.  \left( \sqrt{\frac{32}{27} \left(\frac{M
T_{c}^{0}}{\pi n'}\right)^{3} + 1} - 1 \right)^{1/3}\right]
\frac{T_{c}^{0}}{ 2^{1/3}}   .
\end{eqnarray}
Here $T_{c}^{0}$ is an auxiliary quantity, equal to the
temperature at which the superfluid density vanishes in the
mean-field approximation $n_{s}(T_{c}^{0}) = 0$:
\begin{equation}
\label{tct0} T_c^0 = \left( \frac{2 \pi n' c_s^4 M}{3 \zeta (3)}
\right)^{1/3}  ;
\end{equation}
furthermore, in Eqs.~(\ref{tct}) and (\ref{tct0}),
\begin{eqnarray}
\label{nexx} n' = n  - \frac{n Q}{2Mc_{s}^{2}} .
\end{eqnarray}
The dependence of the Kosterlitz-Thouless transition temperature
$T_{c}$ as a function of the total exciton density $n$ for
different $Q$, obtained from Eq.~(\ref{tct}), is presented in Fig.
1. It can be seen in Fig. 1 that the random field decreases the
Kosterlitz-Thouless transition temperature.  Fig. 2 shows the
dependence of the KTS transition temperature on the random field
parameter $Q$.

This work shows that although the random field depletes the
condensate, Kosterlitz-Thouless superfluidity should still be
possible in a system of spatially indirect excitons. For realistic
experimental parameters
 taken from photoluminescence line broadening measurements\cite{Snoke_paper}, the
 disorder energy is approximately $1$ meV, which implies $Q = 0.05$ in the
 units  of the plots here, which implies a KTS
transition temperature of $T = 12 $ K at a density of $3\times
10^{10}$ cm$^{-2}$. This is still well below the density $n =
1/\rho(D)^2$ at which breakdown of the boson picture of excitons
can occur, which for a typical electron-hole separation of $120$
\AA~, i.e. $D=3$ in the units used here, is approximately $4
\times 10^{11}$ cm$^{-2}$. On the other hand, for larger values of
$Q$, which corresponds to sample quality of just a few years ago,
the critical density can become quite close to this limit.

\section*{Acknowledgements}
O. L.~B. wishes to thank Prof.~Dan~Boyanovsky for many useful and
stimulating discussions. Yu. E.~L. was supported by the INTAS
grant. D. W.~S. and R. D.~C. have been supported by grants from
the National Science Foundation.


\newpage
\newpage

\begin{center}
{\bf Captures to Figures (1-2)}
\end{center}

Fig.1. Dependence of temperature of Kosterlitz-Thouless transition
$T_{c} = T_{c}(n)$ (in units of $Ry* = e^{2}/\epsilon a$; $a =
(4m_{e-h}e^{2}/\epsilon)^{-1}$) on the exciton density $n$ (in
units of $a^{-2}$) at the interwell distance $D = 3$ (in units of
$a$); at different random fields $Q$ (in units of $Ry$): $Q = 0$
-- solid curve; $Q = 0.05$ -- dotted curve; $Q = 0.1$ -- dashed
curve; $Q = 0.2$ -- dashed-dotted curve.

Fig.2. Dependence of temperature of Kosterlitz-Thouless transition
$T_c = T_c (Q)$ (in units of $Ry* = e^{2}/\epsilon a$; $a =
(4m_{e-h}e^{2}/\epsilon)^{-1}$) on the random field  $Q$ (in units
of $Ry$) at the interwell distance $D = 3$ (in units of $Ry$); at
the different exciton densities $n$: $n = 0.005$ -- solid curve;
$n = 0.007$ -- dotted curve; $n = 0.01$ -- dashed curve.

\newpage

\begin{figure}
\rotatebox{270}{
\includegraphics[width = 17cm, height = 16cm]{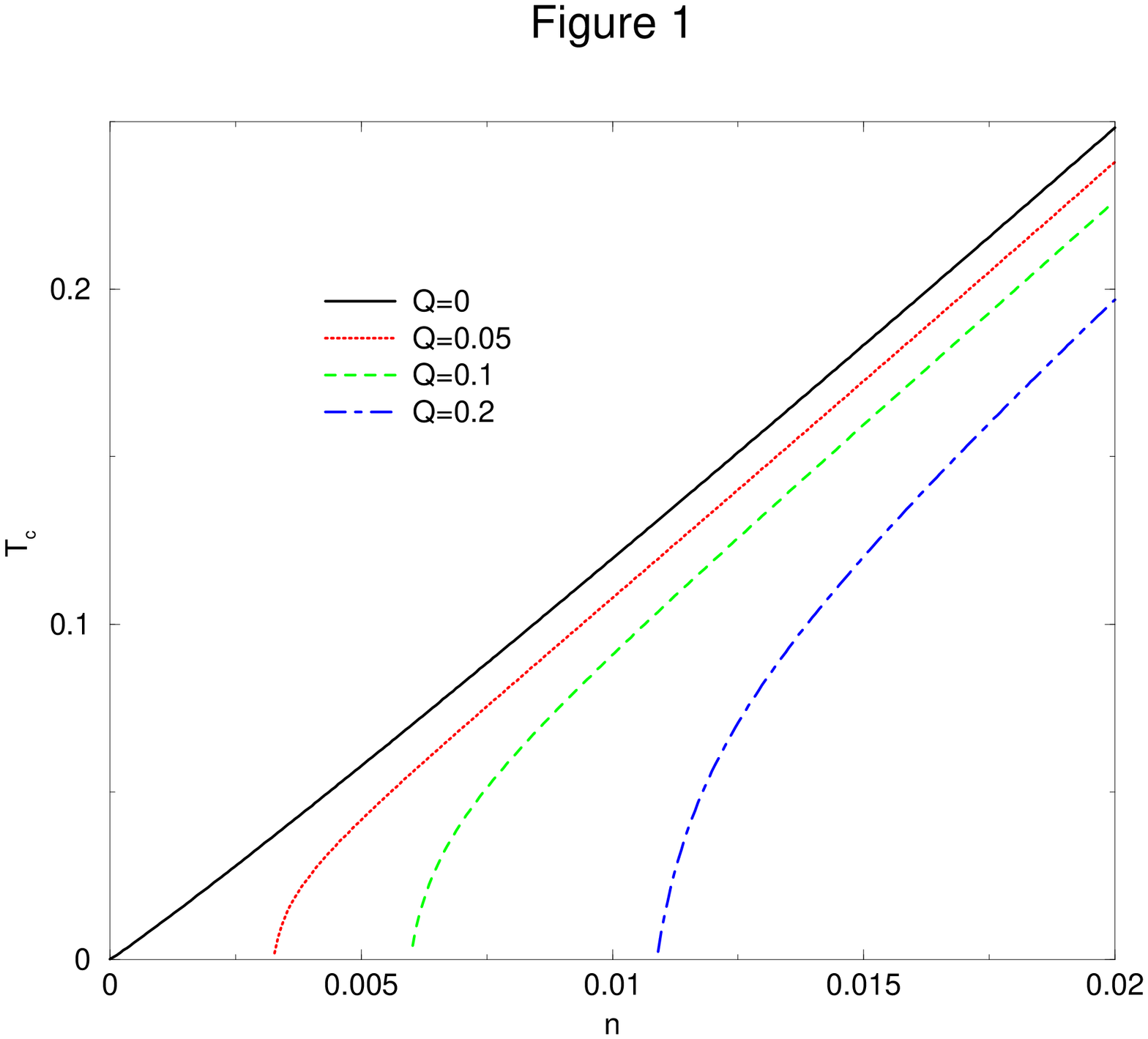}}
\end{figure}

\newpage
\newpage

\begin{figure}
\rotatebox{270}{
\includegraphics[width = 17cm, height = 16cm]{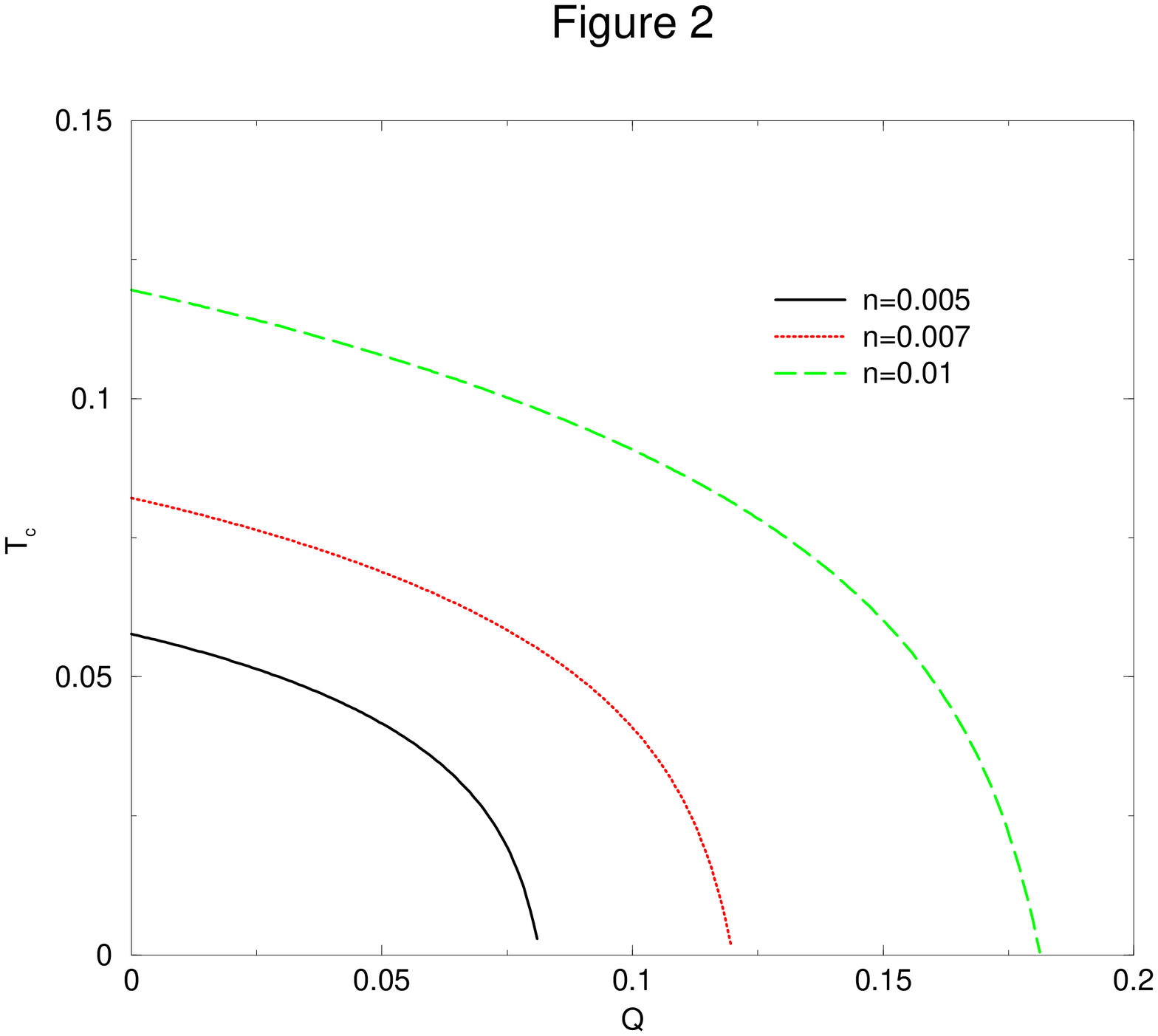}}
\end{figure}

\end{document}